\documentclass[aps,prl,superscriptaddress,twocolumn]{revtex4-1}

\usepackage{amsmath}
\usepackage{amssymb}
\usepackage{mathtools}
\usepackage{graphicx}
\usepackage[usenames,dvipsnames]{color}
\usepackage{bm}
\usepackage{hyperref}
\usepackage{url}
\usepackage[utf8]{inputenc}
\usepackage{subfigure}
\usepackage{slashed,bbm}
\usepackage{graphics,psfrag,epsfig}
\usepackage{dsfont}
\usepackage{setspace}
\usepackage{soul}

\begin{document}

\newcommand{\beq}{\begin{equation}}
\newcommand{\eeq}{\end{equation}}
\newcommand{\beqa}{\begin{eqnarray}}
\newcommand{\eeqa}{\end{eqnarray}}
\newcommand{\note}[1]{{\color{red} [#1]}}
\newcommand{\bra}[1]{\ensuremath{\langle#1|}}
\newcommand{\ket}[1]{\ensuremath{|#1\rangle}}
\newcommand{\bracket}[2]{\ensuremath{\langle#1|#2\rangle}}
\renewcommand{\vec}[1]{\bm{#1}}
\newcommand{\dagga}{{\phantom{\dagger}}}

\newcommand{\txi}{\tilde{\xi}}
\newcommand{\teta}{\tilde{\eta}}

\newcommand{\nn}{\nonumber } 

%-----------------------------------------
\title{\mbox{\hspace{-0.6cm}Excitonic instability and unconventional pairing in the nodal-line materials~ZrSiS~and~ZrSiSe}}

\author{M. M. Scherer}
\affiliation{Institute for Theoretical Physics, University of Cologne, 50937 Cologne, Germany}

\author{C. Honerkamp}
\affiliation{Institut f\"ur Theoretische Festk\"orperphysik,  RWTH Aachen University, and JARA Fundamentals of Future Information Technology, Germany}

\author{A. N. Rudenko}
\affiliation{\mbox{School of Physics and Technology, Wuhan University, Wuhan 430072, China}}
\affiliation{\mbox{Institute for Molecules and Materials, Radboud University, Heijendaalseweg 135, 6525 AJ Nijmegen, The Netherlands}}
\affiliation{\mbox{Theoretical Physics and Applied Mathematics Department, Ural Federal University, Mira Str. 19, 620002 Ekaterinburg, Russia }}

 \author{E. A. Stepanov}
\affiliation{\mbox{Institute for Molecules and Materials, Radboud University, Heijendaalseweg 135, 6525 AJ Nijmegen, The Netherlands}}
\affiliation{\mbox{Theoretical Physics and Applied Mathematics Department, Ural Federal University, Mira Str. 19, 620002 Ekaterinburg, Russia }}

\author{A. I. Lichtenstein}
\affiliation{Institute for Theoretical Physics, University of Hamburg, Jungiusstrasse 9, D-20355 Hamburg, Germany}
\affiliation{\mbox{Theoretical Physics and Applied Mathematics Department, Ural Federal University, Mira Str. 19, 620002 Ekaterinburg, Russia }}

\author{M. I. Katsnelson}
\affiliation{\mbox{Institute for Molecules and Materials, Radboud University, Heijendaalseweg 135, 6525 AJ Nijmegen, The Netherlands}}
\affiliation{\mbox{Theoretical Physics and Applied Mathematics Department, Ural Federal University, Mira Str. 19, 620002 Ekaterinburg, Russia }}

\date{\today}
%-----------------------------------------

%-----------------------------------------
\begin{abstract}
We use a functional renormalization group (fRG) approach to investigate potential interaction-induced instabilities in a two-dimensional model for the Dirac nodal-line materials ZrSiS and ZrSiSe employing model parameters derived from {\it ab initio} calculations.
Our results confirm that the excitonic instability recently found in random-phase approximation for ZrSiS is indeed the leading instability. 
In the simplest modeling, spin- and charge-excitonic states are degenerate.
Beyond this, we show that the fRG analysis produces an energy scale for the onset of the instability in good agreement with the experimentally observed mass enhancement. Additionally, by exploring the parameter space of the model we find that reducing the band splitting increases the instability scale and gives the chance to drive the system into an unconventional superconducting pairing state.
The model parameters for the case of the structurally similar material ZrSiSe suggest the $d$-wave superconducting state as the leading instability with a very small critical scale.
\end{abstract}
%-----------------------------------------

\maketitle

Dirac nodal-line semimetals of the family ZrSiX with X=S, Se or Te, have been investigated intensely in the recent past, in particular because of the unusually large anisotropic magnetoresistance (AMR)\cite{doi:10.1002/aelm.201600228,Alie1601742,Singha2468} and other remarkable effects~\cite{Matusiak2017,doi:10.1021/acs.nanolett.7b02307,PhysRevB.96.195125} that may be rooted in their peculiar electronic structure\cite{PhysRevB.93.201104,Schoop2015,PhysRevB.96.045127}. 
The most studied compound ZrSiS is a layered material consisting of alternating ZrS and Si layers that posses a glide-mirror symmetry with respect to the Si layer. Electronically, node-line semimetals can be viewed as the infinite-degeneracy limit of Dirac or Weyl materials where isolated crossing points of linearly dispersing bands merge to form continuous lines. In ZrSiS, one has closed crossing lines in planes with fixed momemtum $k_z= 0$ or $\pi$ for the direction orthogonal to the layers and also along other high-symmetry lines perpendicular to the $k_x$/$k_y$-plane\cite{Schoop2015}, resulting in a cage-like network of band crossing lines close to the Fermi level. Recent band-structure\cite{2017arXiv171207916R} and photo-emission\cite{ding} works have shown that these band crossings are responsible for the only Fermi surfaces (FS) of the system, which form tubes of varying thickness around the cage-like network of the band crossing lines. In the two-dimensional limit, the FS can be idealized to a tube-like nodal surface formed by two linearly dispersing bands, with diamond-shaped nodal line upon projection into the planes with fixed $k_z$. The quantum Hall effect confirms the Dirac nature of the band structure\cite{Singha2468}.
Another particular feature of ZrSiS is that this linear dispersion spans a wide energy range of up to $\sim 2$eV\cite{Singha2468}. Spin-orbit coupling turns the system into a weak topological insulator\cite{dai}.

Besides the excitement over the electronic structure, quantum oscillation experiments revealed a substantial mass enhancement at temperatures below 10\,K\cite{Pezzini2018}. In a recent theoretical work, Rudenko et al.\cite{2017arXiv171207916R} showed that the basic features of the electronic structure can be understood by a bilayer square-lattice model where the two bands originating in the two layers have opposite signs of the hopping constants and are split in energy by some band splitting $\Delta$. This suffices to generate the diamond-shaped nodal fermion loop in the two-dimensional Brillouin zone (BZ). Then the authors performed a random-phase approximation (RPA) analysis of the effect of intra- and interlayer interactions. This gave clear indication for an excitonic instability at $T \sim 130$\,K. Here, the effective interlayer interaction in RPA diverges, suggesting  the formation of interlayer particle-hole pairs. The self-energy was shown to develop anomalous properties that led to the opening of a pseudogap and a mass enhancement.

In the family ZrSiX, the compound ZrSiSe is structurally similar to ZrSiS and also shows a large AMR that is attributed to a quite similar electronic structure~\cite{2017arXiv170802779P}.

In this work, we employ a functional renormalization group approach (fRG) using the Fermi-surface patching scheme to identify quantum-many body instabilities in an unbiased way. To that end, we use the effective two-orbital Hubbard model on the bilayer square lattice with the set of parameters which approximately reproduces the low-energy effective band structure found in the DFT calculations of Ref.~\onlinecite{2017arXiv171207916R}.
We confirm the previously found excitonic instability and provide additional characterization of it as an onsite spin exciton. Further, we show that the inclusion of all fluctuation channels severely reduces the typical temperatures for the appearance of many-body effects as compared to the ladder approximation, making them compatible with the experimental signature occuring below 10\,K~\cite{Pezzini2018}. 
We also provide a tentative phase diagram of the material with respect to variations of the onsite energy shift and the doping revealing  the vicinity to an unconventional pairing state which might be experimentally accessible. This paired state is actually the leading instability in our model for ZrSiSe, due to a weaker interorbital repulsion based on {\it ab-initio} modeling. Hence, the presence of many-body effects results in quite different ground states as revealed in these two members of the Dirac nodal-line material family that have strong similarity in electronic dispersion but different values of Coulomb interactions.

%-----------------------------fig 1
\begin{figure}[t!]
\includegraphics[height=0.38\columnwidth]{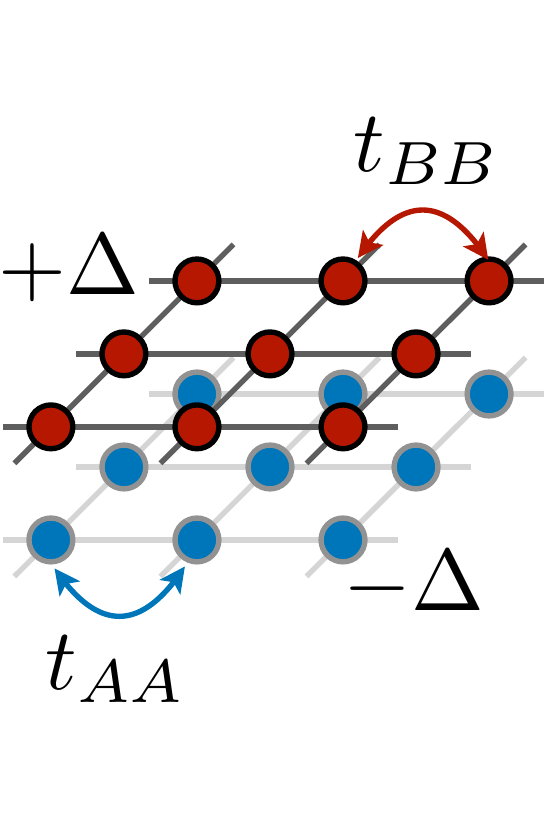}
\hspace{.7cm}
\includegraphics[height=0.38\columnwidth]{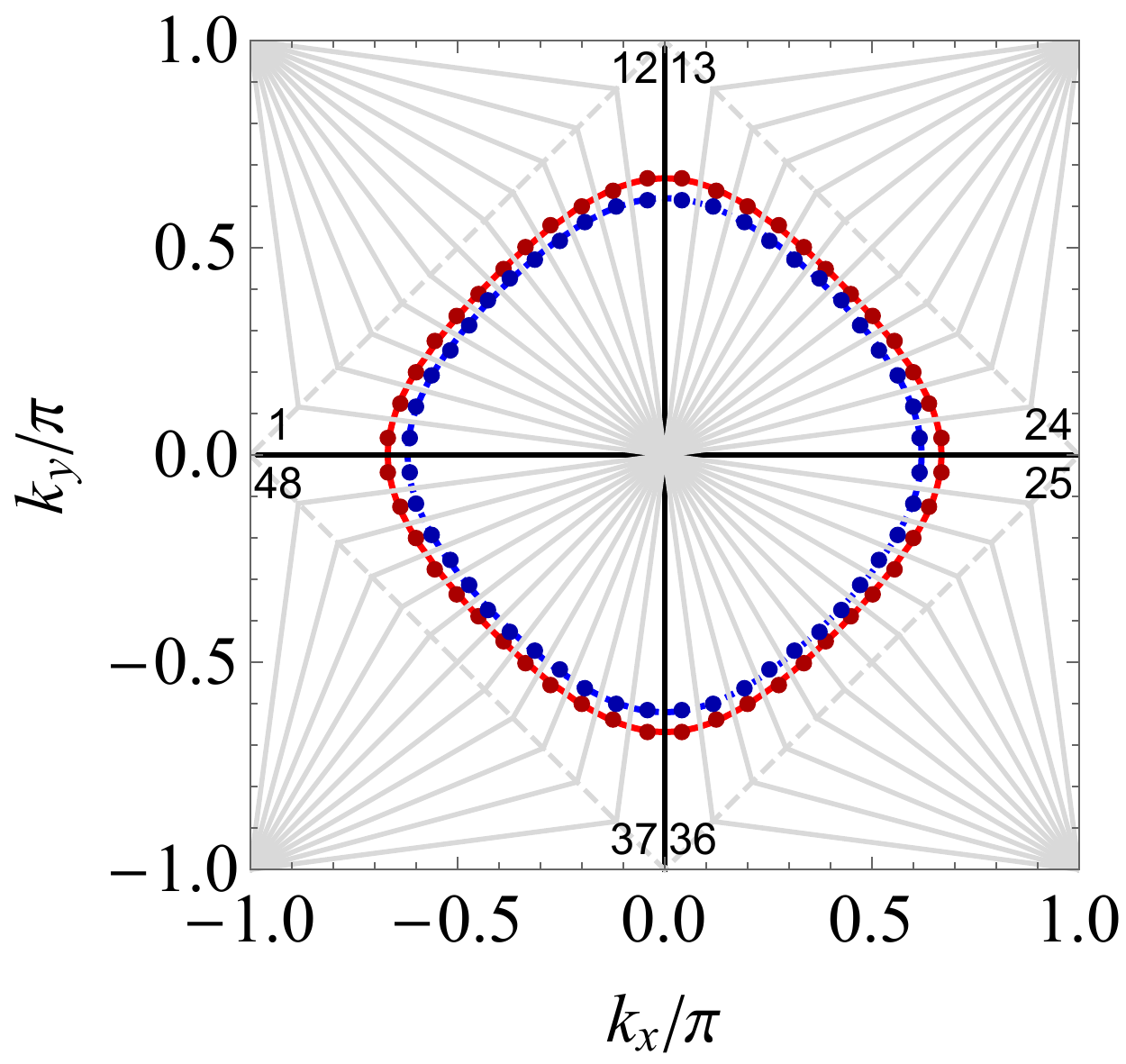}
\caption{(Color online) Left panel: Bilayer square lattice with layers $A$ (bottom) and $B$ (top). The lattice constant is set to $a=1$.
Right panel: Fermi surface patching scheme within the BZ with the tight-binding parameters given in the text. Note that we have added a finite chemical potential of $\mu=0.1$~eV to exhibit the presence of two energy bands.}
\label{fig:latticedisp}
\end{figure}
%-----------------------------

%-----------------------------------------
\paragraph*{Model}
%-----------------------------------------

A low-energy effective tight-binding lattice model reproducing the DFT band structure in the vicinity of the Fermi energy is given by a bilayer square lattice with layers $A,B$ and the single-particle Hamiltonian~\cite{2017arXiv171207916R}
\begin{align}
	H=\hspace{-0.1cm}\sum_{ij \in A}\left(t_{ij}+\Delta\delta_{ij}\right)c_{is}^\dagger c_{js}
	+\hspace{-0.1cm}\sum_{ij \in B}\left(t_{ij}-\Delta\delta_{ij}\right)c_{is}^\dagger c_{js},
	\label{hoppingham}
\end{align}
where we have the electron creation (annihilation) operators $c_{is}^\dagger (c_{is})$, and $s=\uparrow,\downarrow$ is the spin projection. The sums implicitly include summation over the repreated index~$s$. For ZrSiS the nearest-neighbor hoppings between the sites are $t_{AA}\approx -t_{BB}\approx 0.74$\,eV and $\Delta=0.835$\,eV is an onsite energy shift between the layers~\cite{2017arXiv171207916R}. We neglect more remote hopping amplitudes and also the interlayer hopping $t_{AB}\sim 0.1$\,eV, which was found to be significantly smaller than the nearest-neighbor hopping terms~\cite{2017arXiv171207916R}.
This Hamiltonian has two energy bands with dispersion
$
\epsilon_\pm(\vec{k})=\pm 2t_{AA}[\cos k_x+\cos k_y]\mp\Delta-\mu\,.
$ 
The chemical potential $\mu$ allows one to shift the Fermi level.
We show the lattice structure and the band dispersion for $\mu=0$ in Fig.~\ref{fig:latticedisp}. 

This two-dimensional model has a finite density of states (DOS) and electron-hole symmetry, which supports the formation of many-body instabilities. In the three-dimensional Brillouin zone, the FS topology is that of a tube. 
The experimental FS\cite{ding} seems to be closer to a branching network of nodal lines with linearly vanishing density of states, widened into thinner Fermi tubes by additional energy shifts. 
Qualitatively, although a model beyond the leading terms in (\ref{hoppingham}) contains other terms with partially cancelling effects, such a network can be obtained by adding a wavevector-dependent off-diagonal $t_{AB}(\vec{k})$ term.
Nonzero $t_{AB}(\vec{k})$ opens a gap, and where $t_{AB}(\vec{k})$ vanishes linearly in $k_z$, one obtains a linear Dirac DOS. This diminishes the DOS available for the many-body instability. We will discuss this more below and now proceed with $t_{AB}=0$. This seems justified in the context of the observed many-body effects. 

We study first the role of short-range interactions
\begin{align}
H_\mathrm{I}=\sum_{q,\alpha}U_{\alpha\alpha}n_{q,\alpha,\uparrow}n_{-q,\alpha,\downarrow}+\frac{1}{2}\sum_{q,\alpha,\beta}U_{\alpha\beta}n_{q,\alpha}n_{-q,\beta},
\end{align}
with density operator 
$n_{q,\alpha}=\sum_{k}c^\dagger_{k,\alpha}c_{k+q,\alpha},$ where $\alpha \in \{A,B\}$ 
and using the interaction parameters determined in Ref.~\onlinecite{2017arXiv171207916R} for ZrSiS from the constrained RPA, $U_{AA}=2.00$~eV, $U_{BB}=1.22$~eV and $U_{AB}=0.97$~eV.

%-----------------------------------------
\paragraph*{Method}
%-----------------------------------------

%-----------------------------fig 2
\begin{figure}[t!]
\includegraphics[width=\columnwidth]{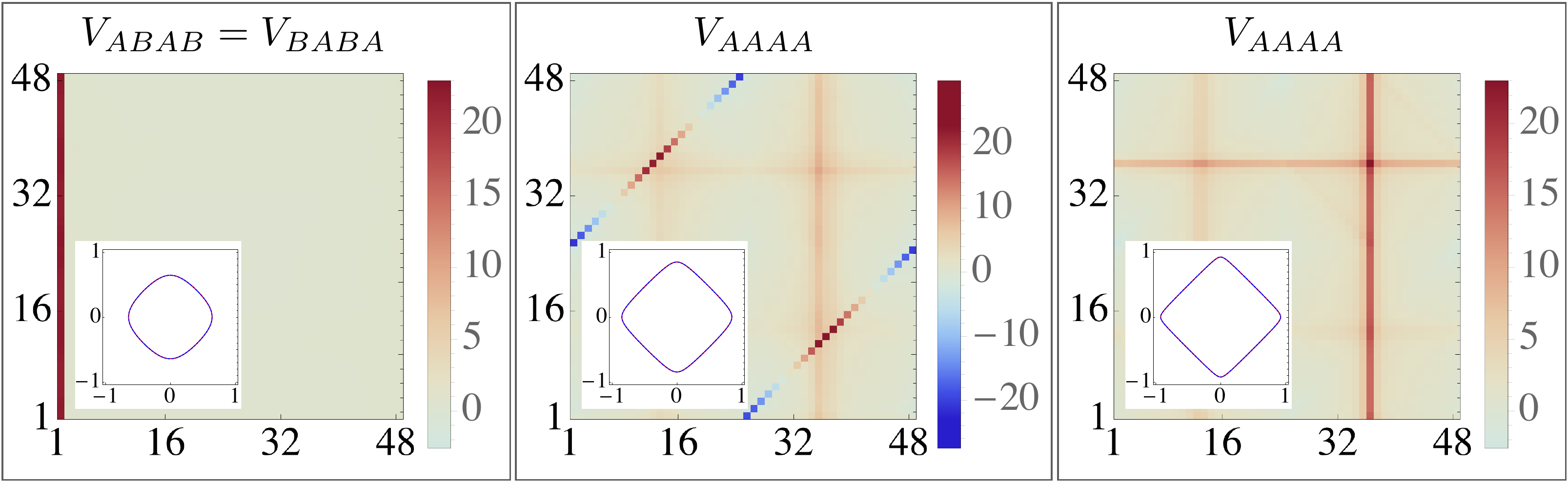}
\caption{(Color online) Fermi surfaces and effective interaction vertex near the critical scale for the appearing instabilities in eV. 
The axes of the vertices $V_{\alpha_1\alpha_2\alpha_3\alpha_4}$, with indicated layer combination, are labelled with the patch number, cf. Fig.~\ref{fig:latticedisp}. Wave vectors $k_1$ are depicted vertically and $k_2$ are depicted horizontally. We fix $k_3$ to be on patch~1. The insets show the Fermi surfaces for the chosen $\Delta/\mathrm{eV}\in \{0.835,0.167,0.04\}$ (from left to right).
Left panel: excitonic instability for $\Delta= 0.835\,\mathrm{eV}$. Middle panel: $d$-wave pairing instability for $\Delta= 0.167\,\mathrm{eV}$. For the choice $\alpha_i=B$ the structure is weaker due to the smaller $U_{BB}$. Right panel: antiferromagnetic instability for $\Delta= 0.04\,\mathrm{eV}$. Again, for the choice $\alpha_i=B$ the structure is weaker due to the smaller $U_{BB}$. The other layer combinations which remain zero in the flow are not shown.}
\label{fig:exciton}
\end{figure}
%-----------------------------

We employ a fRG approach for the one-particle-irreducible vertices of a fermionic many-body system in the Fermi-surface patching scheme~\cite{RevModPhys.84.299,doi:10.1080/00018732.2013.862020,PhysRevB.80.064517,PhysRevB.90.195131}. 
In this approach the four-point function described
by a scale dependent coupling function $V(k_1,k_2; k_3,k_4)$ is renormalized by introducing an infrared cutoff or fRG scale $\Lambda$.
The fRG flow is initialized at the bandwidth and the band structure is integrated out toward the Fermi level by lowering $\Lambda$.
This approximate fRG scheme amounts to an infinite-order summation of one-loop particle-particle and particle-hole terms of second order in the effective interactions.
It allows for an unbiased investigation of the various competing correlations by analysis of the components of $V(k_1,k_2; k_3,k_4)$, which creates instabilities flowing to large values at a fRG scale $\Lambda_c$.
We can use $\Lambda_c$ as an estimate for transition temperatures. 
The discretization of the interaction~$V$ is implemented by dividing the BZ into $N$ patches with constant wavevector dependence within one patch as shown in the right panel of Fig.~\ref{fig:latticedisp}. In addition, the coupling function depends on layer indices.
 
%-----------------------------------------
\paragraph*{Results}
%-----------------------------------------

In Fig.~\ref{fig:exciton} we show the results of the $N$-patch fRG calculations for three different Fermi surfaces, as displayed in the insets. Fig.~\ref{fig:exciton} shows the wavevector structures obtained by the fRG at low scales, where the flow to strong coupling indicative of a ground state change occurs (see, e.g. Ref. \onlinecite{RevModPhys.84.299} for more examples).
Most plots show sharp lines for specific wavevector combinations of the legs of the interaction vertex. These leading features can be expressed as effective long-ranged interactions and hence allow one to determine the fermion bilinears whose expectation values become the order parameter of the potential new ground state. More specifically, in the plots showing the fRG wavevector structures, we have fixed the wavevector $k_3$ to the first patch, cf. Fig.~\ref{fig:latticedisp}, the fourth wavevector is determined by momentum conservation. Fixing $k_3$ on different patches results in a shift of the sharp structures in Fig.~\ref{fig:exciton}. This analysis then allows us to determine an effective reduced form of the interaction vertex $V(k_1,k_2; k_3,k_4)$ where only the sharp structures depending on specific wavevector combinations are taken into account. More details on this procedure can be found in the reviews~\onlinecite{RevModPhys.84.299,doi:10.1080/00018732.2013.862020}.

First, we use the full set of parameters suggested in Ref.~\onlinecite{2017arXiv171207916R}. This produces an instability with the wavevector dependence of the effective interaction shown in the left panel of  Fig.~\ref{fig:exciton}. The wavevector features that flow to strong coupling at the instability can be translated into an effective interaction Hamiltonian reading
\begin{align}
H_{\mathrm{eff}}=-V_0\sum_{i,j}\left(\vec{S}_{i,AB}\cdot\vec{S}_{j,BA}+\frac{n_{i,AB}n_{j,BA}}{4}\right), \label{effint}
\end{align}
where $\vec{S}_{i,AB}=\frac{1}{2}\sum_{s,s'}c^\dagger_{i,A,s}\vec{\sigma}_{ss'}c_{i,B,s'}$ is an onsite interlayer spin operator and $n_{i,AB}$ is a corresponding interlayer bond-density term. This explicit form of the effective interaction vertex near the (excitonic) instability suggests an onsite spin exciton as a possible emergent order parameter. If we fix the spin of this order parameter along the $z$-axis, there will be a non-vanishing long-range ordered expectation value 
$\langle {S}^z_{i,AB} \rangle = \frac{1}{2}\langle c^\dagger_{i,A,\uparrow} c_{i,B,\uparrow} - c^\dagger_{i,A,\downarrow} c_{i,B,\downarrow})$, which lowers the energy of Eq.~\eqref{effint}. 
However, restricting the analysis to the equal-spin fermion bilinears in Eq.~\eqref{effint}, it becomes $H_{\mathrm{eff}}=-\frac{1}{2} V_0\sum_{i,j,s} n_{i,AB,s} n_{j,BA,s} $. From this we see that the coupling between up- and down spins is absent and that the charge exciton with order parameter $\langle \frac{1}{2} n_{i,AB}\rangle = \frac{1}{2} \langle c^\dagger_{i,A,\uparrow} c_{i,B,\uparrow} + c^\dagger_{i,A,\downarrow} c_{i,B,\downarrow}\rangle$ gives the same energy gain. 
The deeper reason for the degeneracy is the $SU(2)_A \times SU(2)_B$ spin rotational symmetry of the model, where the spin frames in the two layers $A$ and $B$ can be rotated freely with respect to each other. This allows us, by flipping the spins in layer $B$, to transform
the bond-spin operator $S^x_{i,AB}= \frac{1}{2} (c^\dagger_{i,A,\uparrow} c_{i,B,\downarrow} + c^\dagger_{i,A,\downarrow} c_{i,B,\uparrow})$ into a bond-density operator $ \frac{1}{2} n_{i,AB}= \frac{1}{2} (c^\dagger_{i,A,\uparrow} c_{i,B,\uparrow} + c^\dagger_{i,A,\downarrow} c_{i,B,\downarrow})$. A nonzero interlayer hopping would lift this degeneracy.

%-----------------------------fig 3
\begin{figure}[t!]
\includegraphics[width=0.9\columnwidth]{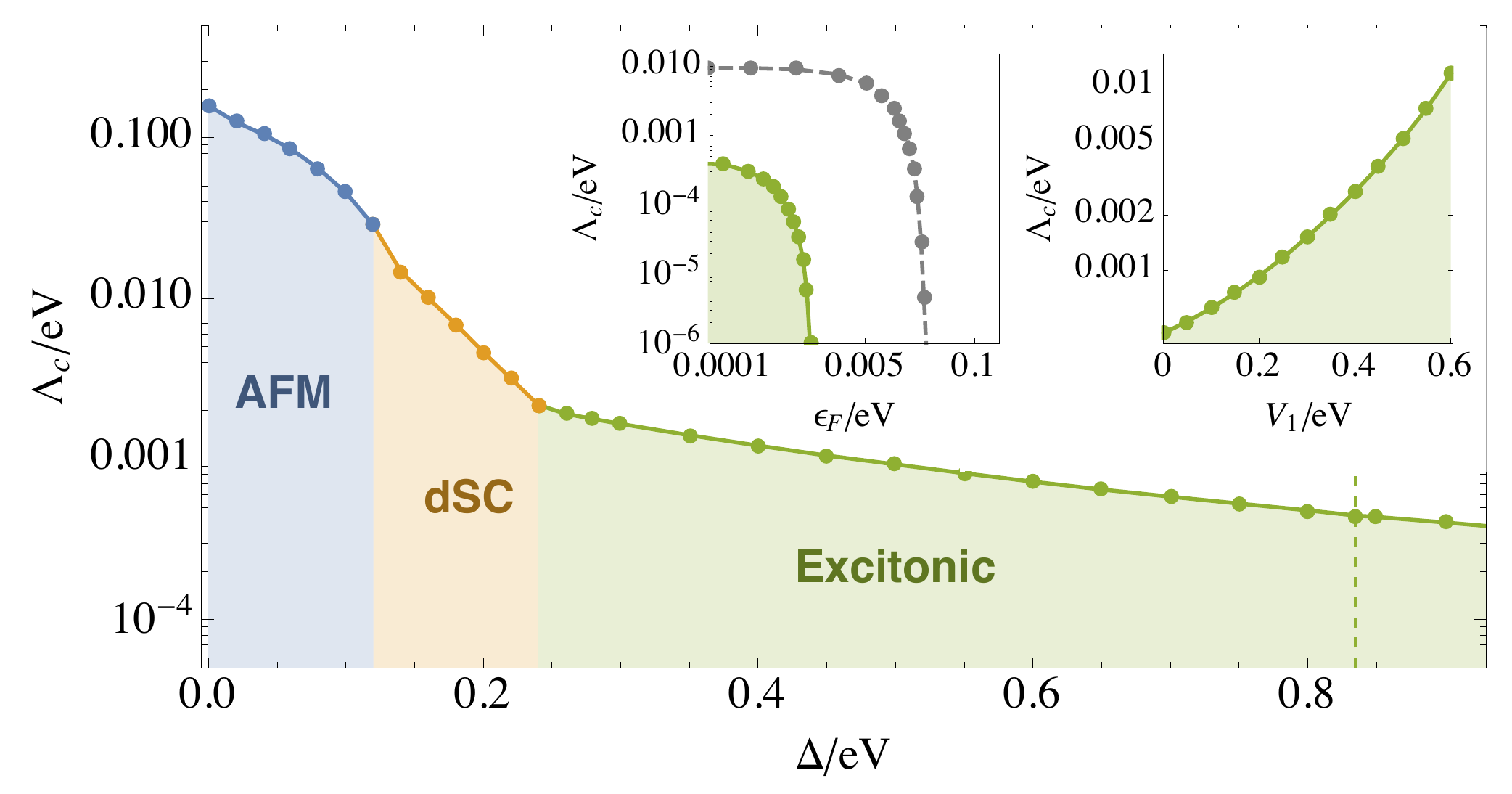}
\caption{(Color online) Tentative phase diagram for the effective model of ZrSiS as a function of the onsite energy shift. The critical scales $\Lambda_c$  can be interpreted as an estimate for the transition temperatures. Left inset: Evolution of $\Lambda_c$ of the excitonic instability with small chemical potentials. The green line shows the full fRG results with all correlation channels active. The gray line corresponds to the crossed particle-hole diagrams, only. Right inset: $\Lambda_c$ of the excitonic instability as a function of nearest-neighbor interaction.}
\label{fig:phasediagram}
\end{figure}
%-----------------------------
%

We find the critical scale of the excitonic instability to be $\Lambda_c\approx 0.47$~meV, which corresponds to a temperature of about 5.5\,K. On first sight, this seems to be in good agreement with the temperatures below which the unconventional mass enhancement is found experimentally~\cite{Pezzini2018}.
We note that this is a significant change as compared to the ladder approximation~\cite{2017arXiv171207916R} which finds a transition temperature of order 100\,K deviating significantly from the experiment.
The difference of these theoretical predictions can be fully attributed to inclusion of all correlation channels within the fRG, an effect that is observed routinely in RG calculations for fermions\cite{katanin2001}. We have explicitly confirmed this by switching off all interaction channels but the particle-hole crossed channel. In this case we find a critical scale consistent with the ordering temperature of the ladder or RPA approximation of Ref.~\onlinecite{2017arXiv171207916R}.
As it was pointed out in Refs.~\onlinecite{dzyaloshinskii1987ie,Bychkov1966,dzyaloshinskii1971,dzyaloshinskii1972} the RPA may provide the incorrect result in
the case of more than one diverging (unstable) scattering channels
presented in the system. For example, it was explicitly shown in Ref.~\onlinecite{katanin2001} that in the Hubbard model near the van Hove filling the Curie temperature is drastically reduced compared to the mean-field value, and
the $d$-wave superconductivity is strongly suppressed by particle-hole
scattering processes. Therefore, the correct description of such problem
requires a simultaneous account for all scattering channels, which is
possible to perform within the parquet or fRG formalism.

Next, we study the dependence of the system and its many-body instabilities on the onsite energy shift parameter $\Delta \in [0,0.835\,\mathrm{eV}]$. We find that depending on the value of $\Delta$ in the given range, there are different types of instabilities. In particular, as $\Delta$ is decreased towards zero, the FS becomes nested reminiscent of the half-filled square lattice at zero doping. In this case an antiferromagnetic (AFM) instability emerges. The AFM instability persists for some finite values of $\Delta$, however, in a small range gives way to a $d$-wave superconducting pairing instability, see the middle panel in Fig.~\ref{fig:exciton}.
We show the dependence on $\Delta$ and the concomitant critical scales $\Lambda_c$ in Fig.~\ref{fig:phasediagram}. Hence, being able to tune the interlayer energy shift $\Delta$ would open the possibility to raise the energy scale of the excitonic instability and to eventually get into a $d$-wave superconducting state with an energy scale $\sim 10$meV for our parameters. 
While the excitonic instability is an interlayer phenomenon, the $d$-wave pairing and AFM instabilities work analogously in the single-layer case with $U_{AB}=0$. The latter get weakened for $\Delta \not= 0$ and give way to the excitonic instability for sufficiently strong $U_{AB}$.

Finite doping or chemical potential $\mu\neq 0$ may destroy the tendency towards the excitonic instability~\cite{2017arXiv171207916R}. In the left inset of Fig.~\ref{fig:phasediagram} we show the critical scale of the excitonic instability as a function of small chemical potential. For comparison, we also show the particle-hole ladder approximation where we only keep crossed particle-hole diagram, corresponding to the approximation in Ref.~\onlinecite{2017arXiv171207916R}. The full fRG with all correlation channels  suppresses the instability scale compared to the ladder study. In both cases, instability range in terms of chemical potential is roughly as large as critical scale $\Lambda_c$ at $\epsilon_F=0$.

Additionally, we study the impact of an intralayer nearest-neighbor density-density repulsion $V_1$, chosen to be identical on both layers, i.e. $V_{1,AA}=V_{1,BB}=V_1\in [0,0.6]$\,eV. Within this range the leading instability still corresponds to the onsite exciton. 
The critical scale shown in the right inset of Fig.~\ref{fig:phasediagram} is strongly increased by $V_1$. This is again an effect beyond the ladder approximation, which would average out the nearest neighbor repulsion in the internal loop summation. The enhancement of the exciton energy scale may be important for understanding the experimental energy scale of a few K. As mentioned above, the low-energy electronic structure of the true material is better described by a line network of Dirac band crossing points than by an open FS with rather flat nonzero DOS as in our 2D model. The Fermi level may not lie directly at the Dirac line, which may also ride up and down in energy. This opens small tubular Fermi surfaces, which again have a finite DOS that is however very likely still smaller than that of the 2D model. Including the diminished density of states near the Fermi level and other model 'imperfections' like the varying Dirac line energy into our theory will likely lower the exciton energy scale. On the other hand, the nonlocal interaction can be expected to be sizable and is found to boost up the exciton scale again. {\it Ab-initio} estimates by the approach used in Ref. \onlinecite{2017arXiv171207916R} produce a value of $V_1 \sim 0.8$\,eV.
While currently we cannot provide detailed fRG calculations on the complex 3D band structure, we identified a mechanism that may be relevant to keep the exciton energy scale at an observable level.

%-----------------------------fig 4
\begin{figure}[t!]
\includegraphics[width=0.85\columnwidth]{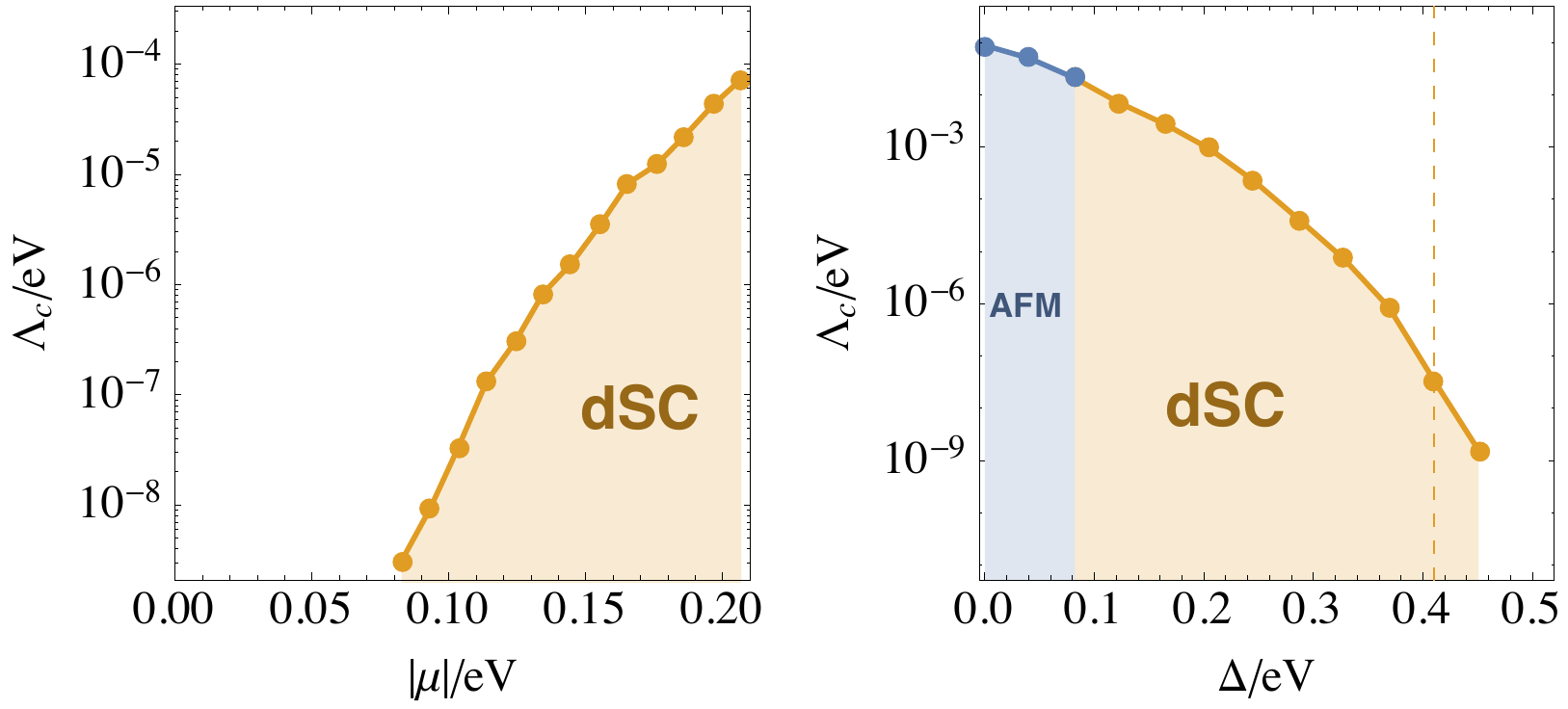}
\caption{(Color online) Left panel: Tentative phase diagram of the effective model for ZrSiSe as a function of chemical potential $\mu$ with fixed onsite energy shift $\Delta=0.41\,\mathrm{eV}$. The critical scales $\Lambda_c$ serve as an estimate for the transition temperatures. For $|\mu|\lesssim 0.08$\,eV no instability can be detected.  
Right panel: $\Lambda_c$ vs. $\Delta$ with $\mu$ adjusted such that the Fermi level lies at the band crossing. The label dSC denotes a $d_{x^2-y^2}$-wave pairing instability, and AFM an antiferromagnetic instability.}
\label{fig:ZrSiSe}
\end{figure}
%-----------------------------
%
Now we turn to the same model with a different set of parameters, appropriate for the  isostructural ZrSiSe. Here, the {\it ab-initio} model building as described in Ref.~\onlinecite{2017arXiv171207916R} gives $t_{AA}=0.4\,\mathrm{eV}, t_{BB}=-0.67\,\mathrm{eV}, \Delta=0.41\,\mathrm{eV}$ for tight-binding parameters and $U_{AA}=1.10\,\mathrm{eV}, U_{BB}=0.82\,\mathrm{eV}$, and a smaller $U_{AB}=0.14\,\mathrm{eV}$ for the interaction parameters. 
The smaller magnitude of these interaction parameters as compared to ZrSiS is related to a more delocalized character of the orbitals in ZrSiSe.
Due to the different nearest-neighbor hopping amplitudes, the Fermi surfaces of the two bands are not congruent at the Fermi level. Here, we adjust the chemical potential accordingly such that at the Fermi level, the Fermi surfaces are identical. We then use the fRG method to identify the leading instability for varying onsite energy shift $\Delta$, see Fig.~\ref{fig:ZrSiSe}. The key difference compared to the ZrSiS case is that, due to the smaller interlayer interaction $U_{AB}$, no excitonic instability occurs. Instead, the physics is closer to the single-layer case. The FS supports the formation of unconventional pairing, which turns out of $d$-wave type (marked as dSC). For very small $\Delta$, the FS is more nested and the instability is toward antiferromagnetic order (denoted as AFM).  The critical scales found in the fRG are quite small, but can at least reach the 0.1meV- or 1K-range when the chemical potential is varied.  
%

%-----------------------------------------
\paragraph*{Conclusions}
%-----------------------------------------

The nodal line semimetal ZrSiS was found to exhibit an anomalous mass enhancement at low temperatures $\lesssim 10$\,K. This was interpreted as the onset of an excitonic instability driven by electronic interactions. 
Here, we performed a theoretical study of the material's quantum many-body instabilities by employing an unbiased functional renormalization group approach to an effective two-dimensional lattice model of the material. 
A related study was recently performed in Ref.~\onlinecite{Moroz2018}.
For the model parameters suggested by ab-initio calculations, we found an interlayer spin-charge degenerate excitonic instability with typical transition temperatures in agreement with the experimental findings. This corroborates the RPA study of Ref.~\onlinecite{2017arXiv171207916R}.
While the density of states near the Fermi level of the three-dimensional material may be smaller than that of our 2D model, other factors like the nearest-neighbor repulsion tend to drastically increase the exciton energy scale.  
Moreover, a variation of the onsite energy shift between the two main electronic states involved in the low-energy electronic structure may induce a transition towards an unconventional superconducting $d_{x^2-y^2}$-pairing state with a gap energy scale $\sim 10$\,meV. This ordering is also found for model parameters appropriate for ZrSiSe, 
with the decisive difference of a less important interlayer repulsion for this compound. Thus, although these nodal-line materials share the same fermiology, differences in the interaction parameters may lead to distinct many-body ground states.

%-----------------------------------------
\paragraph*{Acknowledgments}
%-----------------------------------------

M.M.S. was supported by the DFG through the Collaborative Research Center SFB1238, TP~C04. A.N.R. and E.A.S. acknowledge support from the Russian Science Foundation, Grant No.~17-72-20041. C.H. thanks DFG RTG 1995 for support. A.I.L. acknowledges support from the Cluster of Excellence ``The Hamburg Centre for Ultrafast Imaging (CUI)''
of the German Science Foundation (DFG). The work of M. I. K. was supported by NWO via Spinoza Prize and by ERC Advanced Grant No. 338957 FEMTO/NANO.

%------------------------------------
\bibliography{ZrSiSbib}
%------------------------------------

\end{document}